%% file: paper_sdm.tex
\documentclass[twoside,leqno,twocolumn]{article}
\usepackage{ltexpprt}

\usepackage{booktabs} 
\input{packages}
\usepackage{graphicx}
\usepackage{balance}  

\usepackage{ dsfont }

\begin{document}

\title{Mining Rules Incrementally over Large Knowledge Bases}

\author{Xiaofeng Zhou\thanks{Department of Computer \& Information Science \& Engineering, University of Florida.} \\
\and
Ali Sadeghian\footnotemark[1] \\ {\{xiaofengzhou, asadeghian, daisyw\}@ufl.edu} 
\and
Daisy Zhe Wang\footnotemark[1]}

\date{}

\maketitle


\input{content/abstract.tex}

\input{content/introduction.tex}

\input{content/related_work.tex}

\input{content/preliminaries.tex}
\input{content/incremental_mining.tex}

\input{content/new_metric.tex}
\input{content/experiment.tex}
\input{content/conclusion.tex}

\section*{Acknowledgments}
This work is partially supported by NSF under IIS Award \# 1526753 and DARPA under Award \# FA8750-18-2-0014 (AIDA/GAIA). The authors would also like to thank Anthony Colas, Caleb Bryant and anonymous reviewers for their comments on this paper.

\bibliographystyle{plain}
\bibliography{citation}


\end{document}

%% file: packages.tex
\usepackage{mathtools}
\usepackage{amsmath}

\usepackage{amsthm}
\theoremstyle{definition}
\newtheorem{definition}{Definition}
\usepackage{algorithm}
\usepackage[noend]{algorithmic}
\usepackage{caption}
\usepackage{xcolor}
\usepackage{listings}
\usepackage{xspace}

\usepackage{fixltx2e}
\usepackage{pbox}
\usepackage{enumitem}
\usepackage{arydshln}
\usepackage{adjustbox}
\usepackage{multirow}
\usepackage{subcaption}
\usepackage{upgreek}
\usepackage{amssymb}

\newcommand{\algparbox}[1]{\parbox[t]{\dimexpr\linewidth-\algorithmicindent}{#1\strut}}

\newcommand{\btt}[1]{\texttt{\textbf{#1}}\xspace}
\newcommand{\stt}[1]{\texttt{\small #1}\xspace}
\newcommand{\para}[1]{\vspace{2pt}\noindent\textbf{#1.}\xspace}

\lstdefinelanguage{spark}{
  morekeywords={abstract,aggregateByKey,case,catch,class,def,%
    do,else,extends,false,final,finally,flatMap,%
    for,groupByKey,if,implicit,import,map,match,mixin,%
    new,null,object,override,package,%
    private,protected,reduceByKey,requires,return,saveAsTextFile,sealed,%
    spark,super,textFile,this,throw,trait,true,try,%
    type,val,var,while,with,yield},
  otherkeywords={=>,<-,<\%,<:,>:,\#,@},
  sensitive=true,
  morecomment=[l]{//},
  morecomment=[n]{/*}{*/},
  morestring=[b]",
  morestring=[b]',
  morestring=[b]"""
}

\definecolor{darkblue}{RGB}{0,0,120}
\definecolor{lightgray}{rgb}{0.7,0.7,0.7}

\newcommand{\del}[1]{}
\newcommand{\add}[1]{{#1}}

%% file: content/abstract.tex
\begin{abstract}
Multiple web-scale Knowledge Bases, e.g., Freebase, YAGO, NELL, have been constructed using semi-supervised or unsupervised information extraction techniques and many of them, despite their large sizes, are continuously growing.
Much research effort has been put into mining inference rules from knowledge bases. \del{These rules capture meaningful insights from the data and can be used to deduce missing knowledge in an interpretable fashion.}
To address the task of rule mining over evolving web-scale knowledge bases, we propose a parallel incremental rule mining framework.
%
Our approach is able to efficiently mine rules based on the relational model and apply updates to large knowledge bases; we propose an alternative metric that reduces computation complexity without compromising quality; we apply multiple optimization techniques that reduce runtime by more than 2 orders of magnitude.
Experiments show that our approach efficiently scales to web-scale knowledge bases and saves over 90\% time compared to the state-of-the-art batch rule mining system. We also apply our optimization techniques to the batch rule mining algorithm, reducing runtime by more than half compared to the state-of-the-art. To the best of our knowledge, our incremental rule mining system is the first that handles updates to web-scale knowledge bases.
\end{abstract}


%% file: content/introduction.tex
\section{Introduction}

Knowledge Bases (KBs) have been attracting significant research interest over 
recent years, numerous new KBs are constructed rapidly and old ones growing in size. Examples include YAGO~\cite{hoffart2013yago2},
Google Knowledge Vault~\cite{dong2014knowledge}, and NELL~\cite{carlson2010toward}. These KBs store millions to billions of facts about real-world
entities such as people, places and organizations. Despite their large sizes, KBs are usually growing and 
changing over time. For instance, NELL~\cite{carlson2010toward} has been running continuously since
January 2010 and learning facts from web pages; \del{DBPedia~\cite{auer2007dbpedia}, mined from Wikipedia, maintains a 
live knowledge base whenever a page in Wikipedia changes; }and DeepDive~\cite{shin2015incremental} utilizes database
and machine learning techniques and ``engineer-in-the-loop" development cycles for incremental knowledge base construction.
\\

The growth in size of KBs provides a great opportunity for deducing new information about the world by reasoning over the facts in KBs. While latent feature models such as embedding based models~\cite{toutanova2015representing,sadeghian2016temporal} or methods based on graphical models such as Markov Random Fields~\cite{niu2012elementary} have proven to be effective in inferring new facts from KBs, one major drawback is their lack of interpretability. Another approach is to use graph features based \textit{logical rules}. For example, the rule below
$$liveIn(x, y), isLocatedIn(y, z) \rightarrow liveIn(x, z)$$
can help learn a person's country of residence based on the city where s/he lives. Those rules not only reveal correlations between facts in an explainable way, but also help in various applications including reasoning and expansion over knowledge base~\cite{chen2014knowledge}, \del{cleaning~\cite{nakashole2012query}, }and construction~\cite{carlson2010toward}. Rules are used in other applications such as networks~\cite{jain2018enhanced}, medical geography~\cite{mollalo2018machine} and recommender~systems~\cite{chesnevar2004arguenet}.

While mining rules, specifically mining Horn clauses, have been a major task in Inductive Logic Programming (ILP), classic ILP methods are not scalable for today's large KBs~\cite{chen2016scalekb,galarraga2015fast}s. Amie+~\cite{galarraga2015fast} scales up to 12M facts with pruning and approximation strategies, and OP~\cite{chen2016ontological} adapts Amie+'s mining model and scales to Freebase with 388M facts. However, to the best of our knowledge, no existing work handles dynamic aspect of today's KBs. Those systems work in batches, and the state-of-the-art, OP~\cite{chen2016ontological}, takes 20+ hours to finish on Freebase with fine-tuned parameters, and can only work from scratch if some facts change or new facts are added. At this scale, it is unrealistic to run the batch algorithm again every time a small update to the KB arrives. 

In this paper, we aim to efficiently mine first-order inference rules \textbf{incrementally} over large growing KBs. We propose a novel incremental rule mining framework to mine rules from large evolving KBs by storing the inference rules and facts in relational tables, and using joins \del{to parallelize the searches }to propagate the updates on KBs. \del{We also adopt the pruning techniques from OP~\cite{chen2016ontological} using non-functionality scores in an efficient incremental way, by storing the histogram and using view maintenance techniques~\cite{gupta1993maintaining}.}

However, the intermediate results used to facilitate computing changes in scoring metrics such as support and confidence are still large for efficient update, due to the deduplication needed according to those scoring metrics. Thus we propose two alternative ways to avoid the storage cost: 1) instead of storing intermediate results, we use filter-and-join to remove duplicates before calculating changes to rule metrics; 2) we propose a new metric that avoids the need for deduplication. We compare the runtime of two approaches and compare the new metric with standard confidence. We argue that the new metric performs as good on real knowledge bases and experimentally validate the claim. This new metric saves a significant amount of runtime. 

\del{While our proposed two alternatives try to avoid the storage cost of storing large intermediate result, they still require costly joining of the original KB to identify cases where the update facts appear as head parts in the rules. Thus we apply multiple optimization techniques and reduce the runtime of this operation by nearly 2 orders of magnitude. And we show that our incremental algorithm can handle a 10\% update size to Freebase in less than 1.5 hours, which is less than 10\% runtime of OP, the state of the art batch rule mining system, which requires about 22 hours to handle the update by re-running (to run over the whole KB) again.}

\del{It's also worth mentioning that while OP uses rule-based partitioning by dividing the KB into independent yet overlapping parts, which is a key optimization for OP to scale to Freebase-size KBs, our incremental algorithm does not require this. This simplifies rule mining by avoiding the rule-based partitioning which takes up to 1 hour, and avoiding the storage overhead introduced. 
}
\del{Combining the above algorithms and optimization techniques, }We develop an incremental rule mining system that, to the best of our knowledge, is the first that can handle updates to web-scale KBs efficiently. In summary, the contributions in this paper are:

\begin{itemize}[noitemsep,topsep=2pt,parsep=1pt,partopsep=0pt,leftmargin=10pt,labelindent=0pt,itemindent=0pt]
\item We design an efficient incremental rule mining algorithm that can handle updates to large KBs. 

\item We propose a new metric that avoids searching or storing the prohibitively large intermediate results required in the state-of-the-art systems and show experimentally that the new metric can work as good as competing metrics.
  
\item We apply multiple optimization techniques that avoid unnecessary searching, and handle data skew, which drastically reduces the runtime by nearly 2 orders of magnitude. We also apply these optimization techniques to the state-of-the-art batch rule mining algorithm, OP~\cite{chen2016ontological}, and save half the runtime.

\item We conduct detailed experiments over public large KBs, including YAGO, and Freebase. We validate our optimization techniques, and demonstrate that our system can handle updates to large KBs efficiently.\del{ (i.e., a 10\% update takes less than 10\% of the batch counterpart's runtime). Another benefit of our method is that unlike batch systems, we can output meaningful rules as soon as we finish processing the first few batches of the knowledge base, minimizing the wait time for usable results.}
\end{itemize}


%% file: content/related_work.tex
\section{Related Work}
\label{sec:rel_work}

\noindent\textbf{Mining Horn Clauses.}
Researchers have been working on mining rules from knowledge bases~\cite{ortona2018robust}, especially Horn clauses~\cite{horn1951sentences} since pioneering works on Inductive Logic Programming (ILP)~\cite{muggleton1994inductive}. 
ILP requires
counter examples, which are absent in knowledge bases that implement the open world 
assumption. 
Amie~\cite{galarraga2015fast} extended ILP to handle the absence of  counter examples, and can process KBs in the order of 250K and 12M facts respectively. OP~\cite{chen2016ontological, chen2016scalekb} adopts Amie+'s mining model and by leveraging rule pruning and rule-based partitioning, scales up to Freebase with 388M facts. However, to the best of our knowledge, there is no
existing work that can handle large dynamic knowledge bases. \del{To the best of our knowledge, no existing work is able to mine horn clauses from web-scale knowledge bases that grow dynamically.}

\noindent\textbf{Incremental Association Rule Mining.}
Association rule mining~\cite{agrawal1993mining,fan2015association} is a mature research field relevant to mining Horn rules from KBs. \del{ the rules and facts of a KB to a transactional database. }Various techniques have been proposed for Association rule mining including the Apriori algorithm~\cite{agrawal1993mining}, \del{hashing~\cite{park1995effective}, }partitioning~\cite{savasere1995efficient} and frequent pattern tree~\cite{han2000mining}. Incremental association rule mining has
adapted similar techniques from its batch counterparts~\cite{cheung1996maintenance} for handling new transactions. We use support and 
confidence to measure the significance of rules. However, transactions contribute independently to the itemset counts, while facts in KBs are interconnected by the rules and variables~\cite{chen2016scalekb}. Thus, incremental updates to KBs cannot be trivially updated using techniques from incremental association rule mining.

%% file: content/preliminaries.tex
\section{Preliminaries}
\label{sec:pre}


\subsection{Relational Knowledge Base Model.}

In RDF~\cite{manola2004rdf} knowledge bases, each fact is represented in a triple format (subject, predicate, object). The subject and object refer to real-world entities, and the predicate reveals the relationship between the subject and object. For example, (Barack Obama, LivesIn, USA) encodes the fact that Obama lives in the USA.



\subsection{First-order Rule Mining.}
In KBs such as Freebase~\cite{bollacker2008freebase}\del{ and YAGO~\cite{biega2013inside}}, entities and predicates come from an ontology or schema. Each entity belongs to one or more type and each predicate encodes the relationship between one or more types of entities. For example, the $LiveIn$ predicate specifies the relationship between type person and type place. This information can be used to generate a syntactically correct candidate rule set by traversing the ontology~\cite{chen2016ontological}.

The rules that we aim to mine\del{ from KBs} are of the form:
{\small
\begin{align}
  &(\mathbf{B}\rightarrow H(x,y), ~ \mathbf{w}) \label{eq:horn}
\end{align}
}%
Where the body $\mathbf{B}=\bigwedge_iB_i(\cdot,\cdot)$ is a conjunction of atoms, $H$ is the head predicate, and $\mathbf{w}$ the scoring metric measuring the rule's likelihood.

Similar to OP~\cite{chen2016ontological}, Amie+~\cite{galarraga2015fast} and ILP systems~\cite{muggleton1994inductive},  we focus on mining \textit{closed} and \textit{connected} Horn rules. Two atoms in a rule are connected if they share an entity or a variable, and a rule is connected if each atom is connected transitively to another atom in the rule. A rule is closed if each entity or variable appears at least twice in the rule. The two restrictions are to avoid mining rules with completely unrelated atoms, or rules that predict merely the existence of a fact~\cite{galarraga2015fast}, limiting the search space for rules and retaining the sensitivity of the rules at the cost of expressiveness.

Two first-order clauses are defined to be structurally equivalent if they differ only in predicate symbols but with the same order of variable symbols. For example:
\begin{align*}\small
  &liveIn(x,y),isLocatedIn(y,z) \rightarrow liveIn(x, z)\\
  &actIn(x,y),directedBy(y,z) \rightarrow friendWith(x,z)
\end{align*}
Although the two rules have different predicates and entity types, the structure of the two rules are equivalent in the sense that rule length and order of variables in the rules are the same. This enables us to store the structurally equivalent rules in fixed-column Table~\ref{tbl:equi_rules}, and take advantage of well-optimized relational queries.

\begin{table}[h!]\small
\centering
\begin{tabular}{ |c|c|c| }
 \hline
 \textbf{Body1} & \textbf{Body2} & \textbf{Head} \\ \hline 
 liveIn & isLocatedIn & liveIn\\ \hline
 actedIn & directedBy & friendWith \\ \hline
\end{tabular}
\caption{One table for one type of equivalent rules.}
\label{tbl:equi_rules} 
\end{table}

\subsection{Scoring Metric.}
\label{sec:old_scoring_metrics}
Not all mined rules are interesting or valid. This might be due to missing or erroneous facts in the knowledge base or simply because of inherent exceptions to the rule. Another important factor determining a rule's interestingness is the significance of it, i.e., how many facts can be explained using this rule. To measure the correctness and significance of a rule, different metrics have been proposed in previous related work~\cite{galarraga2015fast, chen2016ontological}, and we briefly review them here. 


\vspace{5pt} \noindent\textbf{Support.} If a rule is only applicable to a small number of facts it is not  that interesting. Support is a measure of significance of a rule. The support of a rule is the number of facts predicted by the rule in the KB.
{\small
\begin{align}
  & \text{supp}(\mathbf{B}\rightarrow H)\coloneqq
  |\{(x,y) |\exists z_1, \dotsc ,z_{n-1} : \mathbf{B}\wedge H(x,y)\}|\label{eq:support}
\end{align}
}%
Where $z_i$'s represent any entities connecting the body.

\vspace{3pt} \noindent\textbf{Confidence.} While support measures the number of correct predictions, confidence defines the quality of the rule based on the ratio of correct predictions to the total number of predictions. The most commonly used confidence is the standard confidence, defined as:
{\small
\begin{align}
  & \text{stdconf}(\mathbf{B}\rightarrow H)\coloneqq
    \frac{\text{supp}(H(x,y)\leftarrow\mathbf{B})}
         {|\{(x,y)|\exists z_2, \dotsc ,z_{n-1} :\mathbf{B}(x,y)\}|} \label{eq:confidence}
\end{align}
}%

The support and standard confidence defined above incur extra computational costs in an incremental rule mining setting, hence, we introduce an alternative confidence in Section~\ref{sec:mining_new_metric} to reduce the workload.

\subsection{Spark Basics.}
Apache Spark~\cite{zaharia2016apache} is a general-purpose cluster computing system that uses the abstraction of resilient distributed datasets (RDD) -- a collection of objects supporting parallel operations. We list a set of operations on RDD used in our method: \add{\stt{map}, \stt{groupByKey}, \stt{reduceByKey} and \stt{join}, further details on Spark website\footnote{https://spark.apache.org/docs/latest/rdd-programming-guide.html}.}
\del{
\begin{itemize}[noitemsep,topsep=2pt,parsep=1pt,partopsep=0pt,leftmargin=10pt,labelindent=0pt,itemindent=0pt]
    \item \stt{map/flatmap} Transform an old RDD to a new RDD by applying a function supplied to each object in the old RDD.
    \item \stt{groupByKey} Group each object in an RDD containing key-value pairs, by the key specified for each object, resulting in a new RDD that has those keys, each mapped to a list of objects with the key.
    \item \stt{reduceByKey} For all objects with the same key in an RDD containing key-value pairs, apply a function to perform a reduce operation on all values of the same key, resulting in a new RDD with the key and resulted values.
    \item \stt{join/leftOuterJoin} Join two RDDs containing key-value pairs, resulting in a new RDD containing new key-value pairs, where the key exists in both RDDs (or left RDD for \stt{leftOuterJoin}), and the value, a tuple containing both values from the two joined RDDs which had the same key.
    The \stt{leftOuterJoin} can also be used to find difference (or filtering) between two large RDDs.
\end{itemize}
}
\del{
We treat facts, both existing facts 
and update facts 
as RDDs containing triples and rely on the parallel operations on RDDs.
}

%% file: content/incremental_mining.tex
\section{Incremental Mining Framework}
\label{sec:inc_mining}

In this section we first give a formal definition of our problem. We then outline our general framework for incremental rule mining, and discuss \del{two}\add{one} commonly used modules in the framework: \del{incremental rule pruning and} incremental inference.

\subsection{Problem Definition.}
\label{sec:prob_def}

Let $\uptau_i ~ (\forall i \in \mathds{N}) $, be a set of an arbitrary number of triples, in the form (subject, predicate, object), conforming with a schema. Let $ \Gamma =  \{\uptau_1, \uptau_2, \uptau_3, \cdots \}$ be an evolving knowledge base and let $\Gamma_c = \{\uptau_1, \uptau_2, \cdots, \uptau_{i-1}\}$ be the currently cumulated KB.  We aim to mine first-order horn clauses from $ \Gamma_c \cup \uptau_i $, given scores previously mined from $ \Gamma_c $.




\subsection{General Framework.}

\begin{figure}[h]\small
\centering
\includegraphics[width=0.44\textwidth]{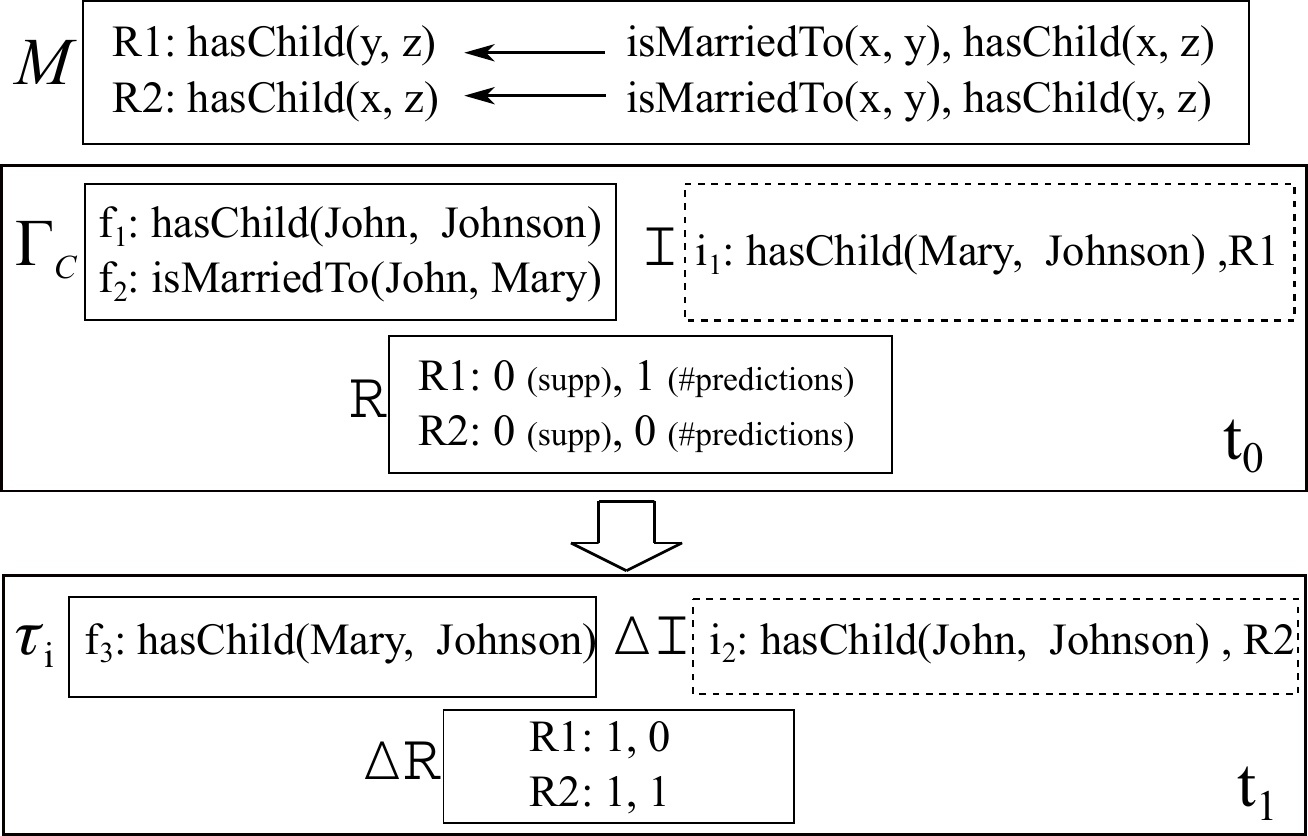}
\caption{Handling an update $\uptau_i$. $M$: candidate rules set, 
$\Gamma_c$: accumulated KB, $I$: intermediate result that can be optionally materialized,
$R$: rules' metrics.}
\label{fig:inc_example}
\end{figure}

The main objective is to compute updated scoring metrics after absorbing new facts $\uptau_i$. Figure~\ref{fig:inc_example} provides a brief example of this update. At time $t_0$, the set of candidate rules $M = \{R1, R2\}$, together with two facts in the existing KB $\Gamma_c$, contributed to the rule metrics $R1: (0, 1)$. We keep the record of the number of predictions instead of the confidence, which can be readily obtained via division of the two values stored.
At time $t_1$, we have an update of fact(s) $\uptau_i$. To update the scoring metrics: 1) the new fact $f_3$ together with fact $f_2$ and rule $R2$ can infer fact $i_2$, which exists in the KB $\Gamma_c$, i.e. a correct prediction, thus contributing $R2: (1, 1)$; 2) the new fact $f_3$ can be inferred from existing facts in $\Gamma_c$, thus contributing $R1: (1, 0)$. By handling the two cases, we obtain the changes to rule metrics $\Delta R$.

Based on the example above, we see that the update facts in $\uptau_i$ can appear in the body, or as the head of a rule, and we need to handle the two cases differently: 
\begin{itemize}[noitemsep,topsep=2pt,parsep=1pt,partopsep=0pt,leftmargin=10pt,labelindent=0pt,itemindent=0pt]
\item Case 1: A fact in $\uptau_i$ appears as a part of the body of some rules. We need to run the inference again by only applying the rules in $M$ (to $\uptau_i \cup \Gamma_c $) that may have a fact in $\uptau_i$ participating in their body. However, inferred facts, or predictions, generated this way might have been inferred before using the same rules with facts from solely $\Gamma_c$. Those duplicate facts inferred by the same rules need to be ruled out when calculating the changes to the metrics. A straightforward way would be to simply store and maintain the previously inferred fact and inference rule id), i.e. the intermediate result ($I$) in Figure~\ref{fig:inc_example}, and subsequently examine if those newly inferred facts exists in $\Gamma_c$ to obtain changes to rule metrics.

\item Case 2: A fact in $\uptau_i$ appears as the head of some rules. We need to update scoring metrics for relevant rules that infer this fact, as their predictions have come true. This is straightforward if we maintain the intermediate result (i.e. predictions with rule IDs in $I$ above): we just check whether the new fact in $\uptau_i$ exists in the intermediate result ($\Delta I$ and $I$); and update scoring metrics for relevant rules.
\end{itemize}

Based on the observation above, we propose Algorithm~\ref{alg:inc-mining}, an incremental rule mining framework which accepts an evolving KB $\Gamma$, with an upper limit for length of rules to be mined\del{, and a functional constraint for rule pruning}.

\begin{algorithm}[ht]\small
  \caption{Inc-Rule-Mining$(\Gamma, max\_len, t)$}
  \begin{algorithmic}[1]
    \STATE $M\leftarrow$ construct-rules($\Gamma$.schema, $max\_len$)\;
    \STATE R $\leftarrow \varnothing$
    \STATE $\Gamma_c \leftarrow \varnothing$
    \FORALL {$(\uptau_i \in \Gamma)$}
      \STATE $\Delta$I $\leftarrow$ Inc-Infer($\Gamma_c$, $\uptau_i$, $M$)\;
      \STATE R $\leftarrow$ R $\cup$ Infer-Update($\Delta$I, $\Gamma_c$) \;
      \STATE $\Gamma_c\leftarrow\Gamma_c\cup\uptau_i$\;
      \STATE R $\leftarrow$ R $\cup$ Check-Update($\uptau_i$)\;
      \STATE \textbf{output} R \;
    \ENDFOR
  \end{algorithmic}
  
  \label{alg:inc-mining}
\end{algorithm}

Algorithm~\ref{alg:inc-mining} outlines the major steps of our incremental framework: first we generate the syntactically valid rules from the schema (Line 1) using the ontological path-finding algorithm~\cite{chen2016ontological}, then we process each update $\uptau_i$ in $\Gamma$ iteratively. In each iteration, we run incremental inference to get inferred facts with rule IDs (Line 5),  to get the changes to the scoring metrics for case 1 where the update facts $\uptau_i$ act as body of rules (Line 6). Finally we get the changes to metrics for case 2 where update facts appears as head of rules (Line 8).

We focus on the commonly used module \del{of Inc-Prune and }Inc-Infer in the rest of this section. Calculating the changes to scoring metrics in Infer-Update and Check-Update will be detailed in Section~\ref{sec:mining_w_std} and Section~\ref{sec:mining_new_metric}.

\subsection{Incremental Inference.}
Before calculating the changes to candidate rules' scoring metrics, we need to infer new facts with the rule IDs in $\Delta I$ using the update facts in $\uptau_i$. 
Without loss of generality, we describe our algorithms based on the following class of rules:

\stt{\hspace{20pt} h(x, y) <- b\textsubscript{1}(x, z), b\textsubscript{2}(z, y)}

\vspace{2pt}
\noindent It is straightforward to generalize to other type of rules (see~\cite{chen2016ontological} for the full list of rule types).

\begin{algorithm}[ht]\small
  \caption{\textsc{Inc-Infer}($\Gamma_c, \uptau_i, M)$}
  \begin{algorithmic}[1]
   \REQUIRE $\Gamma_c$ = \{(\stt{pred\textsubscript{c},sub\textsubscript{c},obj\textsubscript{c}})\}
   \REQUIRE $\uptau_i$ = \{(\stt{pred\textsubscript{i},sub\textsubscript{i},obj\textsubscript{i}})\}
   \REQUIRE $M$ = \{(\stt{ID,head,body1,body2})\}
    \STATE \stt{preds} $\leftarrow$ $M$.groupBy(\stt{body1})
    \STATE \btt{FlatMap} {each fact (\stt{pred\textsubscript{i}, sub\textsubscript{i}, obj\textsubscript{i}}) $\in$ $\uptau_i$ to: 
     for all (\stt{ID, head, body1, body2}) $\in$ \stt{preds}.get(\stt{pred\textsubscript{i}})\\
     yield list of \stt{((body2,obj\textsubscript{i}), (head,sub\textsubscript{i},ID))}} \\
     
    \STATE \btt{Map} \algparbox{each fact \stt{(pred\textsubscript{c},sub\textsubscript{c},obj\textsubscript{c})} $\in$ $\Gamma_c$ \\
               to \stt{((pred\textsubscript{c},obj\textsubscript{c}), sub\textsubscript{c})} } \\
               
    \STATE \btt{Join} \algparbox{output from Line 2 and Line 3 on \stt{(body2,obj\textsubscript{i})}\\
      = \stt{(pred\textsubscript{c},obj\textsubscript{c})}
       to get \\  
       \{\stt{((pred\textsubscript{c},obj\textsubscript{c}), ((head,sub\textsubscript{i},ID),sub\textsubscript{c}))}\}   } \\
     
    \STATE \stt{res1} $\leftarrow$ \btt{Map} \algparbox{ each \\ \stt{((pred\textsubscript{c},obj\textsubscript{c}),((head,sub\textsubscript{i},ID),sub\textsubscript{c}))} \\
    to  \stt{(head,sub\textsubscript{i},sub\textsubscript{c},ID)}  } 
     
    \STATE \stt{res2} $\leftarrow$ \COMMENT{similarly for $\uptau_i$ appears as \stt{body2} in $M$}
    \STATE \stt{res3} $\leftarrow$ \COMMENT{similarly for $\uptau_i$ appears as both \stt{body1} and \stt{body2} in $M$}
    \RETURN \stt{res1} $\cup$ \stt{res2} $\cup$ \stt{res3}
  \end{algorithmic}
  \label{alg:inc-infer}
\end{algorithm}

Algorithm~\ref{alg:inc-infer} infers new predictions given new facts $\uptau_i$, cumulated KB $\Gamma_c$ and candidate rules set $M$. There are 3 cases to handle in Algorithm~\ref{alg:inc-infer}, the new facts in $\uptau_i$ only act as $body1$, or $body2$, or both. For brevity, we only show how to handle the first case and the rest can be similarly implemented. First we do a map-side join on $\uptau_i$ with rules in $M$ (Line 1-2), then we join the result with $\Gamma_c$, which participate as $body2$ (Line 3-4), to get the inferred facts with rule IDs (Line 5). We start from the incremental update $\uptau_i$, which we assume should be much smaller than $\Gamma_c$, thus avoiding generating large intermediate results as much as possible.

\del{\noindent\textbf{Complexity Analysis}
The runtime complexity of Algorithm~\ref{alg:inc-infer} is dominated by the intermediate result size of join operation in Line 4. To analyze the size of intermediate result $\Delta I$, we start from Line 2, which is a map-side join of $M$ and $\uptau_i$, thus its size is at-most $|M||\uptau_i|$. In Line 3, suppose the maximum size of each key (\stt{pred\textsubscript{c},obj\textsubscript{c}}) is $c_{max}$, then the maximum join size in Line 4 is $c_{max}|M||\uptau_i|$. 
}

\del{Theoretically $c_{max}$ can be as large as $|\Gamma_c|$. As for real-world KBs, due to the highly skewed power-law degree distributions of real-world sparse graphs~\cite{gonzalez2012powergraph}, $c_{max}$ can actually be very large: in YAGO2s and Freebase, $c_{max}$ can be as large as $0.176|\Gamma_c|$ = 780875 and $0.029|\Gamma_c|$ respectively. However, the mean number of occurrences of (pred, obj) pair is merely $5.8$ and $1.9$. Assuming the output from Line 4 follows similar distribution as that from Line 3, a more realistic estimate would be the weighted average
number of each (pred, obj) pair $\bar{c}$. Thus the expected complexity of Algorithm~\ref{alg:inc-infer} would be $O(\bar{c}|M||\uptau_i|)$. For YAGO2s and Freebase, $\bar{c}=738$ and $1493$ respectively. 
}

\section{Mining Using Standard Confidence.}
\label{sec:mining_w_std}

Here we first introduce our vanilla approach where the intermediate result $I$ is
stored and maintained. Then we propose a searching approach to avoid maintaining the large intermediate result. The searching approach requires calculating partial $I$ on the fly by searching the whole KB $\Gamma_c$. We discuss how to optimize this operation. We also improve the performance of batch rule mining algorithms using those optimization techniques.

\subsection{Vanilla Approach.}
Maintaining a deduplicated copy of intermediate result $I$ makes updating facts $\uptau_i$ straightforward. First, obtain the deduplicated $\Delta I$ via Algorithm~\ref{alg:inc-infer}: for case 1, remove tuples in $\Delta I \cap I$ from $\Delta I$, and join with $\Gamma_c$ to calculate the changes for the relevant rules' metrics; for case 2, update facts $\uptau_i$ can be compared with $I \cup \Delta I$ to obtain changes for relevant rules. For brevity, we omit the algorithms here.

\subsection{Mining via Searching.}
To avoid storing the intermediate result $I$, we can run the inference again on $\Gamma_c$ to obtain $I$ on the fly. However, the complexity would be equivalent to the batch mining algorithm. To minimize the inference cost, we filter $\Gamma_c$ before inference. 

\begin{algorithm}[h]\small
  \caption{\textsc{Infer-Update\textsubscript{Search}}($\Delta I, \Gamma_c, M)$}
  \begin{algorithmic}[1]
   \REQUIRE $\Gamma_c$ = \{(\stt{pred\textsubscript{c},sub\textsubscript{c},obj\textsubscript{c}})\}
   \REQUIRE $\Delta I$ = \{(\stt{head,sub,obj,ID})\}
   \REQUIRE $M$ = \{(\stt{ID,head,body1,body2})\}
   \STATE $\Delta I$ $\leftarrow$ Distinct $\Delta I$
   \STATE \btt{leftOuterJoin} filter out tuples in $\Delta I$ that also exists in Search($\Delta I$, $\Gamma_c$, $M$)
   \STATE \btt{leftOuterJoin} remaining $\Delta I$ with $\Gamma_c$, yield \stt{(ID, (exist,1))}, where 
   \stt{exist=1} if the inferred fact exists, otherwise \stt{exist=0}.
   \STATE \btt{reduceByKey} summing the \stt{(exist, 1)} pairs for each \stt{ID}
  \end{algorithmic}
  \label{alg:infer-update-v2}
\end{algorithm}

\begin{algorithm}[h]\small
  \caption{\textsc{Check-Update\textsubscript{Search}}($\uptau_i, \Gamma_c, M)$}
  \begin{algorithmic}[1]
   \REQUIRE $\uptau_i$ = \{(\stt{head\textsubscript{i},sub\textsubscript{i},obj\textsubscript{i}})\}
   \REQUIRE $\Gamma_c$ = \{(\stt{pred\textsubscript{c},sub\textsubscript{c},obj\textsubscript{c}})\}
   \REQUIRE $M$ = \{(\stt{ID,head,body1,body2})\}
   \STATE \stt{tmp} $\leftarrow$ \btt{Distinct} Search($\uptau_i$, $\Gamma_c$, M)
   \STATE \btt{Join} $\uptau_i$ with tmp, and issue \stt{(ID, (1,0))} tuples.
   \STATE \btt{reduceByKey} summing the \stt{(1, 0)} pairs for each \stt{ID}
  \end{algorithmic}
  \label{alg:check-update-v2}
\end{algorithm}

Algorithm~\ref{alg:infer-update-v2} deals with case 1 where the update facts $\uptau_i$ appear in the body of the rules. Instead of maintaining intermediate result $I$, we generate partial $I$ that includes $\Delta I$ (i.e. Algorithm~\ref{alg:search} tries to infer new facts from $\Gamma_c$ with rules $M$ that can possibly `hit' facts in $\Delta I$ for deduplicating $\Delta I$).

Algorithm~\ref{alg:check-update-v2} deals with case 2 similarly, but instead of using the intermediate result $I$, we generate the partial incremental result that could possibly infer $\uptau_i$ from existing facts $\Gamma_c$ on the fly. This is done using Algorithm~\ref{alg:search} detailed below.

\subsubsection{Searching.}
Algorithm~\ref{alg:search} enables us to search the large intermediate results without materializing it, thus significantly improving performance. First we filter $\Gamma_b$ with the entities (subjects and objects) in $\Gamma_s$, to get $f1$ and $f2$ and group them by $obj$ to get the list of $(pred, sub)$ pairs for each $obj$. Finally, we join the grouped lists on $obj$, and for each matched pair of lists connected by $obj$, we apply join Algorithm~\ref{alg:group-join-adaptive} to obtain the inferred facts, i.e. the partial $I$ relevant to input $\Gamma_s$

\begin{algorithm}[ht]\small
  \caption{\textsc{Search}($\Gamma_s, \Gamma_b, M)$}
  \begin{algorithmic}[1]
   \REQUIRE $\Gamma_s$ = \{(\stt{pred\textsubscript{s},sub\textsubscript{s},obj\textsubscript{s}})\}
   \REQUIRE $\Gamma_b$ = \{(\stt{pred\textsubscript{b},sub\textsubscript{b},obj\textsubscript{b}})\}
   \REQUIRE $M$ = \{(\stt{ID,head,body1,body2})\}
   \STATE \stt{f1} $\leftarrow$ \btt{filter} $\Gamma_b$ that have \stt{sub\textsubscript{b}} appearing as \stt{sub\textsubscript{s}} in $\Gamma_s$
   \STATE \stt{f2} $\leftarrow$ \btt{filter} $\Gamma_b$ that have \stt{obj\textsubscript{b}} appearing as \stt{obj\textsubscript{s}} in $\Gamma_s$
   \STATE \stt{g1} $\leftarrow$ \btt{groupByKey} \stt{f1} yielding list of \stt{(pred,sub)} for each \stt{obj}, with max group size limit \stt{m}.
   
   \STATE \stt{g2} $\leftarrow$ \btt{groupByKey} \stt{f2} yielding list of \stt{(pred,obj)} for each \stt{sub}, with max group size limit \stt{m}.
   \STATE \btt{Join} \stt{g1} with \stt{g2} (apply Algorithm~\ref{alg:group-join-adaptive}) yielding tuples of \stt{(pred,sub,obj,ID)}
  \end{algorithmic}
  \label{alg:search}
\end{algorithm}

Algorithm~\ref{alg:group-join-adaptive} is a hash-join algorithm. We process each list of $(pred, sub)$ pairs by searching for rules in $M$ that can be applied to those pairs, and output inferred facts and rule IDs if available.
The algorithm minimizes the search cost by iterating through the smaller list or rules set and searching the larger via hash-maps.

\begin{algorithm}[ht]\small
  \caption{\textsc{Group-Join-Adaptive}(\stt{obj}, \newline
       $l1$=\{(\stt{pred, sub})\},  $l2$=\{(\stt{pred, sub})\},  $M$)}
  \begin{algorithmic}[1]
    \IF{$M.size < l1.size * l2.size$} 
       \STATE \stt{preds1} $\leftarrow$ $l1$.groupBy(\stt{pred})
       \STATE \stt{preds2} $\leftarrow$ $l2$.groupBy(\stt{pred})
       \FOR{$r \in M$} 
          \FOR{\stt{sub1} $\in$ \stt{preds1}.get(r.\stt{body1})}
             \FOR{\stt{sub2} $\in$ \stt{preds2}.get(r.\stt{body2})}
               \STATE emit \stt{(r.head,sub1,sub2,r.ID)}
             \ENDFOR
          \ENDFOR
       \ENDFOR
    \ELSE 
      \STATE \stt{rules} $\leftarrow$ M.groupby((\stt{body1, body2}))
      \FOR{(\stt{pred1, sub1}) $\in$ $l1$}
         \FOR{(\stt{pred2, sub2}) $\in$ $l2$}
            \FOR{r $\in$ \stt{rules}.get((\stt{pred1, pred2}))}
               \STATE emit \stt{(r.head,sub1,sub2,r.ID)}
            \ENDFOR
         \ENDFOR
      \ENDFOR
    \ENDIF
  \end{algorithmic}
  \label{alg:group-join-adaptive}
\end{algorithm}

\del{\noindent\textbf{Complexity Analysis}
The complexity of the searching approach is dominated by Algorithm~\ref{alg:infer-update-v2}, which checks whether $\Delta I$ can be inferred from $\Gamma_c$ or not using Algorithm~\ref{alg:search}. And the complexity of algorithm~\ref{alg:search} is $O(t|M|d|\Gamma_s|)$ , assuming that the largest degree a object/subject can get is $d_{max}$, thus after filtering the size of \stt{f1} and \stt{f2} is $O(d_{max}|\Gamma_s|)$. Similar to the analysis of Algorithm~\ref{alg:inc-infer}, the $d_{max}$ can be very large due to the skewed distribution of the natural graphs, we use the weighted average $\bar{d}$ to estimate the expected complexity. For real KBs like YAGO2s and Freebase, $(\bar{d}_{sub}, \bar{d}_{obj})$ is $(389, 1042)$ and $(2393, 2426)$ respectively. Thus the expected complexity of Algorithm~\ref{alg:search} is $O(\bar{d}t|M||\Gamma_s|)$.
Since the size of $\Delta I$ is $O(\bar{c}|M||\uptau_i|)$, the complexity is 
$O(t|M|\bar{d}|\Gamma_s|) = O(t|M|\bar{d}\bar{c}|M||\uptau_i|) = O(t\bar{d}\bar{c}|M|^2|\uptau_i|)$. 
}

\subsubsection{Optimization Techniques.}
\label{subsec:optimization}
While algorithm~\ref{alg:search} is similar to batch rule mining algorithm in OP~\cite{chen2016ontological} for applying the rules to generate the intermediate result, there are three important improvements\del{ in our algorithm}.

\begin{itemize}[noitemsep,topsep=2pt,parsep=1pt,partopsep=0pt,leftmargin=10pt,labelindent=0pt,itemindent=0pt]
\item Adaptive Join:  Algorithm~\ref{alg:group-join-adaptive} tries to minimize the cost for searching matching pairs. Due to the skewed power-law degree distribution of natural knowledge graphs, most of the matched lists are very small, a few of the matched lists can be very large. Different from OP where the searching loop only starts from rules set $M$, which usually contains a significant number of candidate rules, our algorithm starts searching from the smaller one, $M$ or matched lists $l1$ and $l2$.

\item Handle Data Skew: The runtime of Algorithm~\ref{alg:group-join-adaptive} is dominated by the largest matched lists in Line 5. This data skew causes single or a few long running tasks and leads to resource under-utilization. We impose a group size limit for each group in Algorithm~\ref{alg:search} (Line 3-4) to re-distribute the workload evenly. \del{This handles the data skew effectively.}

\item Adaptive Filter: In Algorithm~\ref{alg:search} there are two approach to filter: 1) broadcast the subjects/objects of $\Gamma_s$ to each worker to apply filtering; 2) join the subjects/objects of $\Gamma_s$ with $\Gamma_b$ to apply filtering. While the broadcasting method is only suitable to small updates, large updates mandate join-filtering method. We apply an adaptive filtering approach to minimize the filtering cost\del{, reducing the runtime of algorithm~\ref{alg:search} by up to 10\% - 30\%}.
\end{itemize}

In Section~\ref{sec:experiment} we conduct experimental analysis and show that our optimization techniques above can reduce runtime drastically by almost 2 orders of magnitude.

%% file: content/new_metric.tex
\section{Mining Using New Metric}
\label{sec:mining_new_metric}
The two incremental mining algorithms in the previous section either require storing the large intermediate result or suffer from large search burden. This
is due to standard scoring metrics being holistic. In the definition of support~(\ref{eq:support}) and confidence~(\ref{eq:confidence}), the number of distinct $H(x, y)$ pairs are counted. Thus we need to deduplicate the intermediate result, which becomes costly in the incremental setting.

\del{Here we propose a new confidence metric that avoids the deduplication need and the third version of our incremental algorithm.}

\subsection{New Confidence Metric}



We introduce a new metric for measuring the correctness of a rule: \textit{xconf}. Our metric measures confidence by normalizing the different number of instantiations with the total number of body-only instantiations: 
{\small
\begin{align}
  & \text{xconf}(\mathbf{B}\rightarrow H)\coloneqq
    \frac{ |\{ (z_1, \dotsc ,z_{n}) | H(z_1,z_n)\leftarrow\mathbf{B} \}| }
         { |\{ (z_1, \dotsc ,z_{n}) | \mathbf{B}\}|} \label{eq:xconfidence}
\end{align}
}%
In other terms, x-confidence does not force a uniqueness constraint as in standard confidence. \del{This metric is very similar to the confidence metric used in the well-studied problem of association rule mining~\cite{agrawal1994fast}.}  

While using the rule's number of instantiations to measure \textit{support} is not generally suitable, the number of instantiations is not monotonic, i.e., with this definition adding new atoms to a rule can artificially increase the support~\cite{galarraga2015fast}. For example, consider the rule:
\begin{center}\small
    $BornIn(x,y) \leftarrow LivesIn(x,y)$
\end{center}
adding the extra condition $LivesIn(z,y)$ to the rule's body makes a narrower rule, but the support increases for every~\stt{z}. 

Nevertheless, it does not imply that the number of instantiations cannot be used to measure \textit{confidence}. In \textit{xconf} the denominator also incorporates the instantiations, and this normalizes the effect of adding new atoms mentioned above, except in very few corner cases. \del{In fact, as we will see in experimental Section~\ref{sec:experiment}, rules mined from two widely used knowledge bases using this metric have a \textit{x-confidence} very close to the standard confidence.}
Additionally, rules longer than length 3 mostly reduce to shorter rules, or are erroneous, and are expensive to learn as runtime grows exponentially as rule length grows~\cite{chen2016ontological}. Thus we are mostly interested in closed horn rules of length 2 or 3. In such rules, the problems mentioned above do not arise and one can use the number of instantiations to measure significance.

It is trivial to support deletion of facts using new metric once addition is implemented, as no duplication is needed. As for standard confidence scores, supporting deletion requires keeping track of path counts of the body in instantiations for the same head, resulting in slight increase in storage cost and runtime. For simplicity we omit the deletion in this paper.

\subsection{Mining Using Xconf}

Algorithm~\ref{alg:infer-update-v3} is adapted from Algorithm~\ref{alg:infer-update-v2} to handle the first case where the update facts $\uptau_i$ appear in the body of rules. Here we do not deduplicate $\Delta I$ and avoid filtering out duplicate tuples in $\Delta I$ using `Search' algorithm.

\begin{algorithm}[ht]\small
  \caption{\textsc{Infer-Update\textsubscript{xconf}}($\Delta I, \Gamma_c)$}
  \begin{algorithmic}[1]
   \REQUIRE $\Gamma_c$ = \{$(pred_c, sub_c, obj_c)$\}
   \REQUIRE $\Delta I$ = ${(head, sub, obj, rule\_ID)}$
   \STATE \btt{leftOuterJoin} $\Delta I$ with $\Gamma_c$, yield $(rule\_ID, (exist, 1))$, where 
   exist=1 if the inferred fact exists, otherwise exists=0.
   \STATE res $\leftarrow$ \btt{reduceByKey} summing the $(exist, 1)$ pairs for each $rule\_ID$
  \end{algorithmic}
  \label{alg:infer-update-v3}
\end{algorithm}

Algorithm~\ref{alg:check-update-v3} is almost identical to Algorithm~\ref{alg:check-update-v2}, except that we remove the deduplicate operation and directly calculate the changes to new metric with the possible inferred facts from 'Search', as all the body instantiations contribute separately.

\begin{algorithm}[ht]\small
  \caption{\textsc{Check-Update\textsubscript{xconf}}($\uptau_i, \Gamma_c, M)$}
  \begin{algorithmic}[1]
   \REQUIRE $\uptau_i$ = ${(head_i, sub_i, obj_i)}$
   \REQUIRE $\Gamma_c$ = \{$(pred_c, sub_c, obj_c)$\}
   \REQUIRE $M$ = \{(\stt{ID,head,body1,body2})\}
   \STATE tmp $\leftarrow$  Search($\uptau_i$, $\Gamma_c$, $M$)
   \STATE \btt{Join} $\uptau_i$ with tmp, and issue $(rule\_ID, (1, 0))$ tuples.
   \STATE \btt{reduceByKey} summing the $(exist, 1)$ pairs for each $rule\_ID$
  \end{algorithmic}
  \label{alg:check-update-v3}
\end{algorithm}

\del{\noindent\textbf{Complexity Analysis} Similar to previous analysis, the runtime complexity of Algorithm~\ref{alg:infer-update-v3} is $\bar{c}|M||\uptau_i|$, while algorithm~\ref{alg:check-update-v3} has complexity $\bar{d}\bar{t}|M||\uptau_i|$. Thus the overall expected complexity of our incremental rule mining using \texttt{xconf} is $O(\bar{d}\bar{t}|M||\uptau_i|)$.
}


%% file: content/experiment.tex
\section{Experiment}
\label{sec:experiment}

We conduct experiments on two real-world KBs: YAGO~\cite{hoffart2013yago2} and Freebase~\cite{bollacker2008freebase}. First, we compare the 3 variants of our incremental algorithm with the state-of- the-art batch rule mining system OP~\cite{chen2016ontological}.  We show that for different update sizes, our incremental algorithms can easily save more than 90\% of the time, compared to re-running OP again. Second, we compare our new metric with the standard confidence in terms of rule quality and confirm that our new metric is close to standard confidence scores on real KBs. Third, we show how our optimization techniques for algorithm~\ref{alg:search} speed up our incremental algorithm.

\del{We begin by describing the datasets and our experiment setup.}

\para{YAGO}
YAGO~\cite{hoffart2013yago2} is a multilingual knowledge base derived from Wikipedia, WordNet and GeoNames. We use the same version of YAGO2s used in OP.

\para{Freebase}
Freebase~\cite{bollacker2008freebase} is a large collaborative knowledge base constructed from many sources. Freebase has 1.9B facts\footnote{https://developers.google.com/freebase/} and we preprocess the dataset by removing the multi-language support and use the remaining 344M facts. Statistics of the KBs are listed in Table~\ref{tab:datasets}

\begin{table}[ht]\footnotesize
  \setlength{\tabcolsep}{2pt}
  \centerline{
  \begin{tabular}{|c|p{31mm}|}\hline
    \textbf{KB} & \multicolumn{1}{c|}{\textbf{Size}}\\ \hline
    \textbf{YAGO2s} & \pbox[c][.72cm][c]{3cm}{\# Entities: 2,137,468\\
                                              \# Facts: 4,484,907} \\\hline
    \textbf{Freebase} & \# Entities: 110,459,875 \newline \# Facts: 344,192,734 \\\hline
  \end{tabular}~~~%
  \hspace{3pt}
  \begin{tabular}{|c|c|}\hline
    \textbf{Max length} & 3 \\\hline
    \textbf{Max Group Size} & 30,000 \\ \hline
    \textbf{Min support} & 0 \\\hline
    \textbf{Min confidence} & 0.0 \\\hline
  \end{tabular}}
  \caption{(Left) Datasets statistics. (Right) Default parameters.}
  \label{tab:datasets}
\end{table}

\para{Experimental Setup}
We run experiments on a 64-core machine running AMD Opteron Processor (6376), with 512GB memory and 3TB disk space on Ubuntu 14.04.3 LTS with kernel version 3.13.0-68-generic, with software: Spark 2.2.0, in Scala 2.12 and Java 1.8. 

\subsection{Incremental vs Batch.}

\del{\textbf{Baseline: Batch Systems}} We use the state-of-the-art batch rule mining system OP~\cite{chen2016ontological} as the baseline. We also include the runtime of our implementation on batch rule mining enhanced with our optimization techniques~\footnote{ https://bitbucket.org/datasci/fast-rule-mining}. Results are summarized in Table~\ref{tab:batch_runtime}: OP takes 24.3 mins and 21.9 hours to finish mining on YAGO2s and Freebase, respectively. Our batch implementation takes 11.5 mins and 9.0 hours, less than half the time of OP.


\begin{table}[h!]\footnotesize
  \centering
  \begin{tabular}{|l|c|r|}\hline
    Runtime   &  YAGO2s    & Freebase     \\ \hline
    OP        &  24.3 mins & 21.9 hrs     \\ \hline
    Fast OP\textsuperscript{1} (ours) &  11.5 mins & 9.0 hrs      \\ \hline
  \end{tabular}
  \caption{Batch Runtime}
  \label{tab:batch_runtime}
\end{table}

To evaluate the incremental algorithms, we randomly divide both datasets into two parts: 90\% as the base ($\Gamma_c$), and updates with different sizes($\uptau_i$) from 1\% to 10\%. We evaluate the incremental algorithms by applying updates of different size to the base. We report the runtime of our incremental algorithms.

\begin{figure}[h]
\centering
\begin{subfigure}[b]{.32\textwidth}
  \includegraphics[width=1\linewidth]{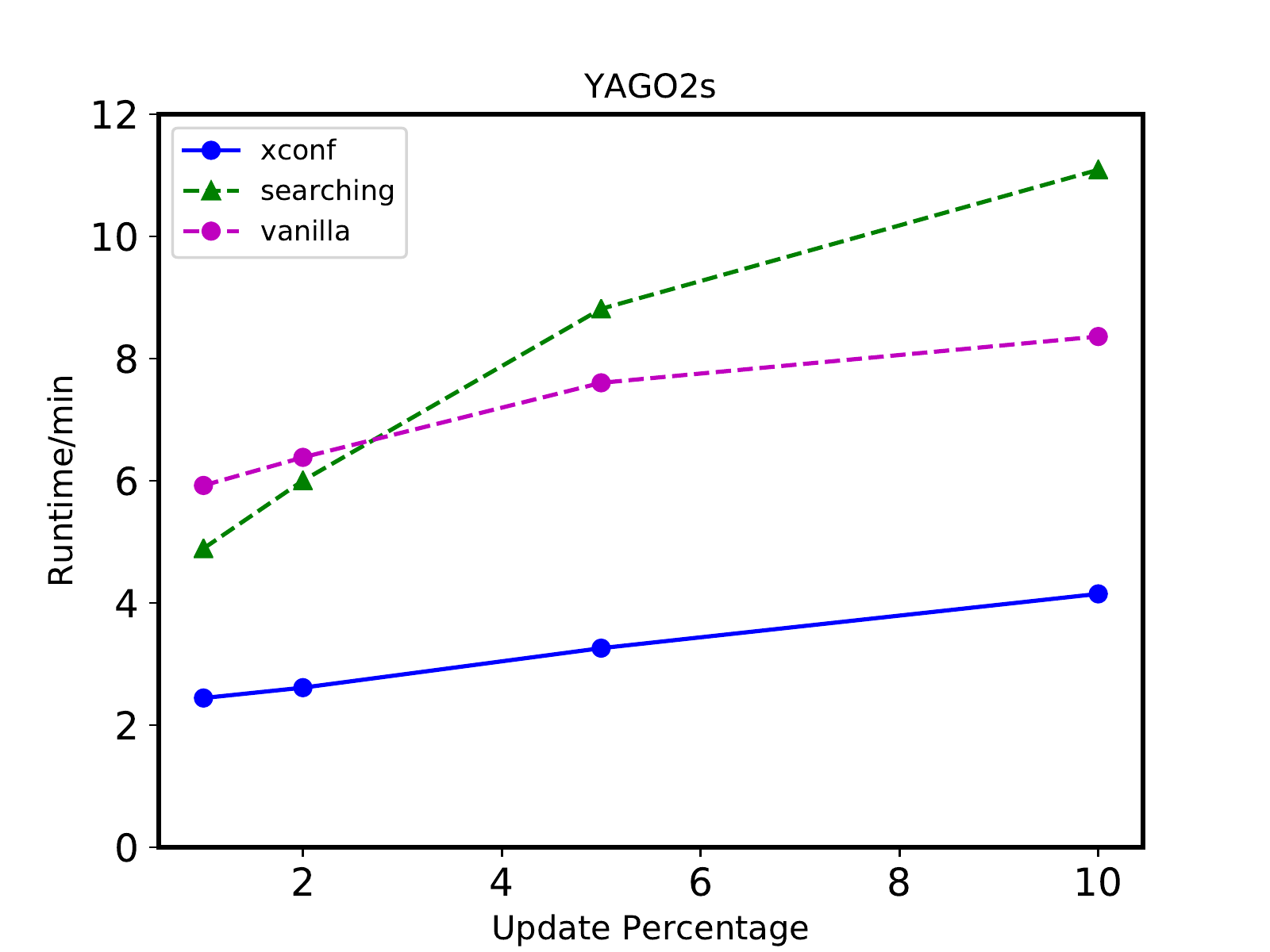}
  \caption{}
  \label{fig:yago2s_diff_update}
\end{subfigure}
\begin{subfigure}[b]{.32\textwidth}
  \includegraphics[width=1\linewidth]{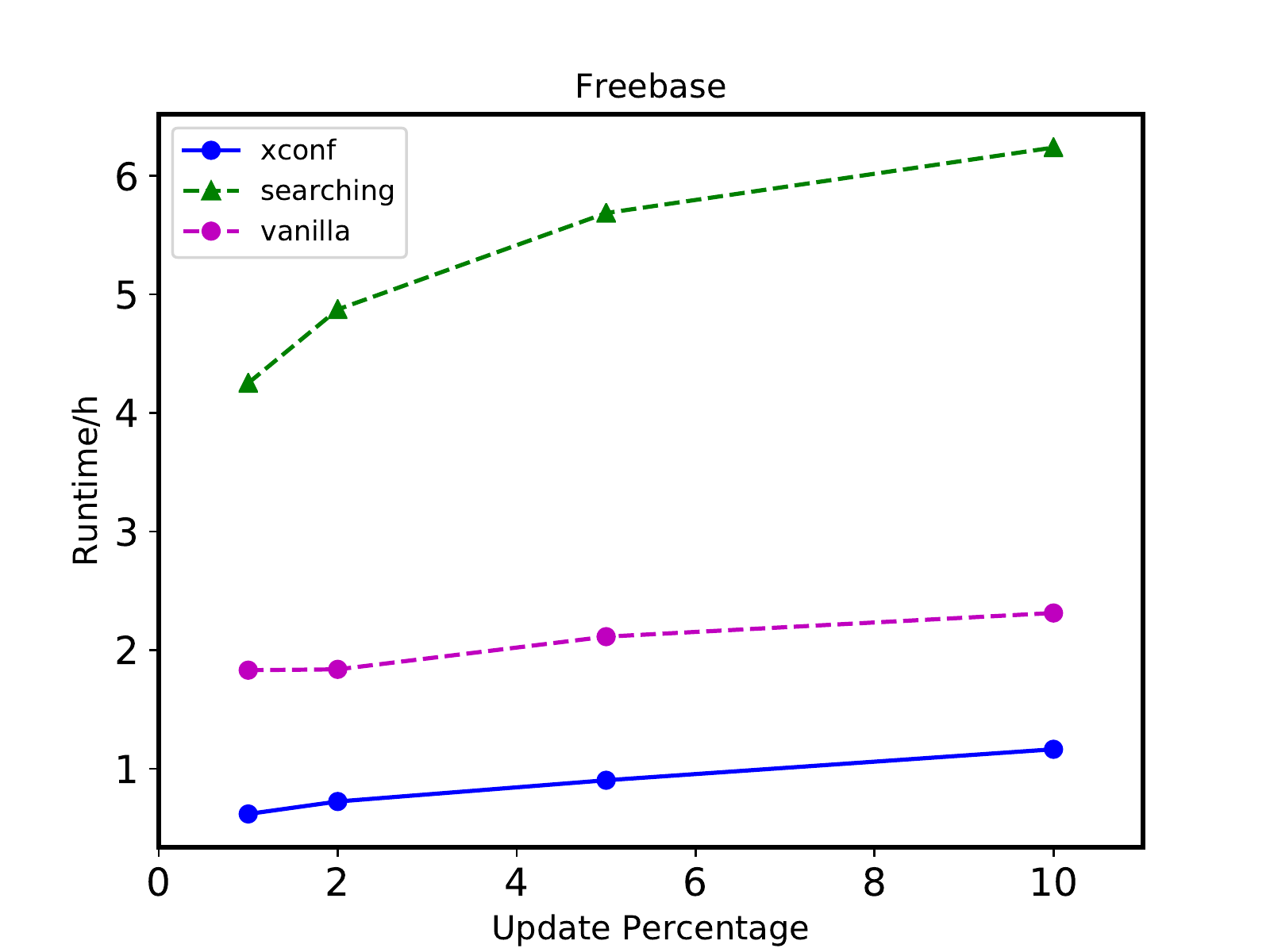}
  \caption{}
  \label{fig:freebase_diff_update}
\end{subfigure}
\caption{runtime with different update size}
\end{figure}

Figure~\ref{fig:yago2s_diff_update} shows the runtime of our incremental algorithms on YAGO2s with different update sizes. The incremental version with the new metric 'xconf' takes about 2.4 mins for 1\% update and 4.1 mins for 10\% update. While for the vanilla approach, the runtime is 5.9 mins for a 1\% update and 8.3 mins for 10\% update; and for the searching approach, the runtime goes up from 4.9 mins to 11.1 mins for 1\% update and 10\% update, respectively.
\del{The longer runtime for the vanilla approach is due to the large intermediate result with 201M records, about 45 times larger than YAGO2s. This also explains the small increase in runtime from 1\% update to 10\% update.}

Similarly Figure~\ref{fig:freebase_diff_update} shows the incremental mining runtime on Freebase. 
The new metric xconf performs the best: for 1\% update size, it only takes about 0.61 hours to apply the update and get the changes to the rule scoring metric, and only 1.16 hours to apply a 10\% update, which accounts for almost less than 95\% of the batch counterpart. The vanilla approach can handle 1\% within 1.8 hours and 10\% in 2.4 hours, at the cost of storing intermediate result more than 10 times larger than Freebase which can grow rapidly as the KBs get less sparse. As for the searching approach, the runtime for 1\% update takes about 4.2 hours and 10\% update takes about 6.2 hours, about 6 times slower than the new metric and 2-3 times slower than the vanilla approach. Still, the searching approach takes about an acceptable 25\% of  OP's time to handle 10\% update size. The performance boost shows that our optimization techniques are more effective for larger KBs.



\subsection{Xconf vs. Stdconf}

\begin{figure*}[bht!]
\centering
\includegraphics[width=0.80\textwidth]{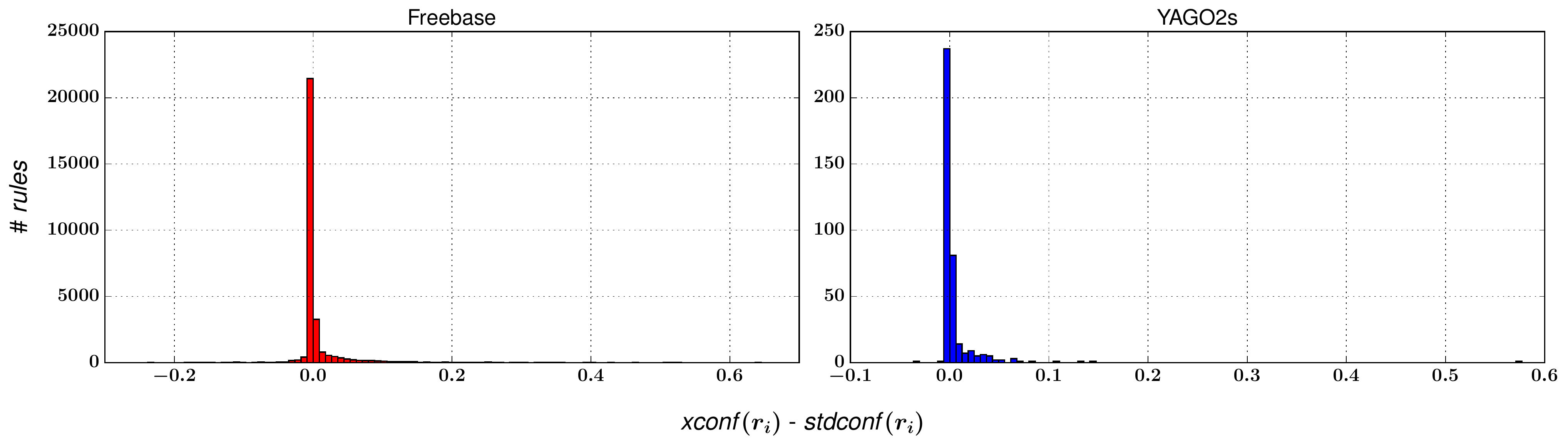}
\caption{Histogram of difference between the two confidence metrics of rules mined on YAGO2s and Freebase.}
\label{fig:xconf_stdconf_hist}
\end{figure*}

We compare \textit{xconf} to \textit{stdconf} via the rules mined from YAGO2s and Freebase\del{ using the two metrics respectively}. From the definitions in~(\ref{eq:confidence})~and~(\ref{eq:xconfidence}), it is clear that a rule will have a non-zero \textit{stdconf} iff it has a non-zero \textit{xconf}, thus both algorithms will produce the same set of rules but possibly with different confidence values. The histogram of difference between the two metrics for each rule on both datasets is shown in Figure~\ref{fig:xconf_stdconf_hist}. 
Almost all rules mined have a very small difference in confidence score when measured with the two metrics. When different, \textit{xconf} is usually higher than \textit{stdconf} in both KBs.
Only a few rules have a significant difference in confidence scores, and some are shown in Table~\ref{tbl:diff_rules}. 

\begin{table*}[bht!]\footnotesize
\centering

\begin{tabular}{llcc}
\multicolumn{2}{c}{Rule}                                       & \textit{stdconf} & \textit{xconf} \\ \hline
\multirow{3}{*}{YAGO2s} &  dealsWith(z,x) $\wedge$ imports(z,y)  $\rightarrow$ exports(x,y)                & 0.06      & 0.13    \\
                        &  influences(z,x) $\wedge$ hasGender(z,y)   $\rightarrow$ hasGender(x,y)           & 0.81      & 0.89    \\
                        &  dealsWith(z,x) $\wedge$ dealsWith(y,z)   $\rightarrow$ dealsWith(x,y)            & 0.31      & 0.45   \\ \hdashline[4pt/5pt]
\multirow{2}{*}{Freebase}   & \shortstack[r]{ model.parent\_aircraft\_model(x,z) $\wedge$ model.manufacturer(z,y) $\rightarrow$ model.manufacturer(x,y)} & 0.68 & 0.74 \\
                    &  \shortstack[r]{location.partially\_contains(x,z) $\wedge$ location.containedby(z,y) $\rightarrow$ location.partially\_contains(x,y)} & 0.06 & 0.10 \\
\end{tabular}
\caption{Link prediction results for various rule mining methods}
\label{tbl:diff_rules}
\end{table*}

\subsection{Effect of Optimization Techniques}
We experimentally examine how optimization techniques in Section~\ref{subsec:optimization} reduce runtime by 2 orders of magnitude.

\begin{figure*}[!htb]
\centering
\minipage{0.30\textwidth}
  \includegraphics[width=1.0\textwidth]{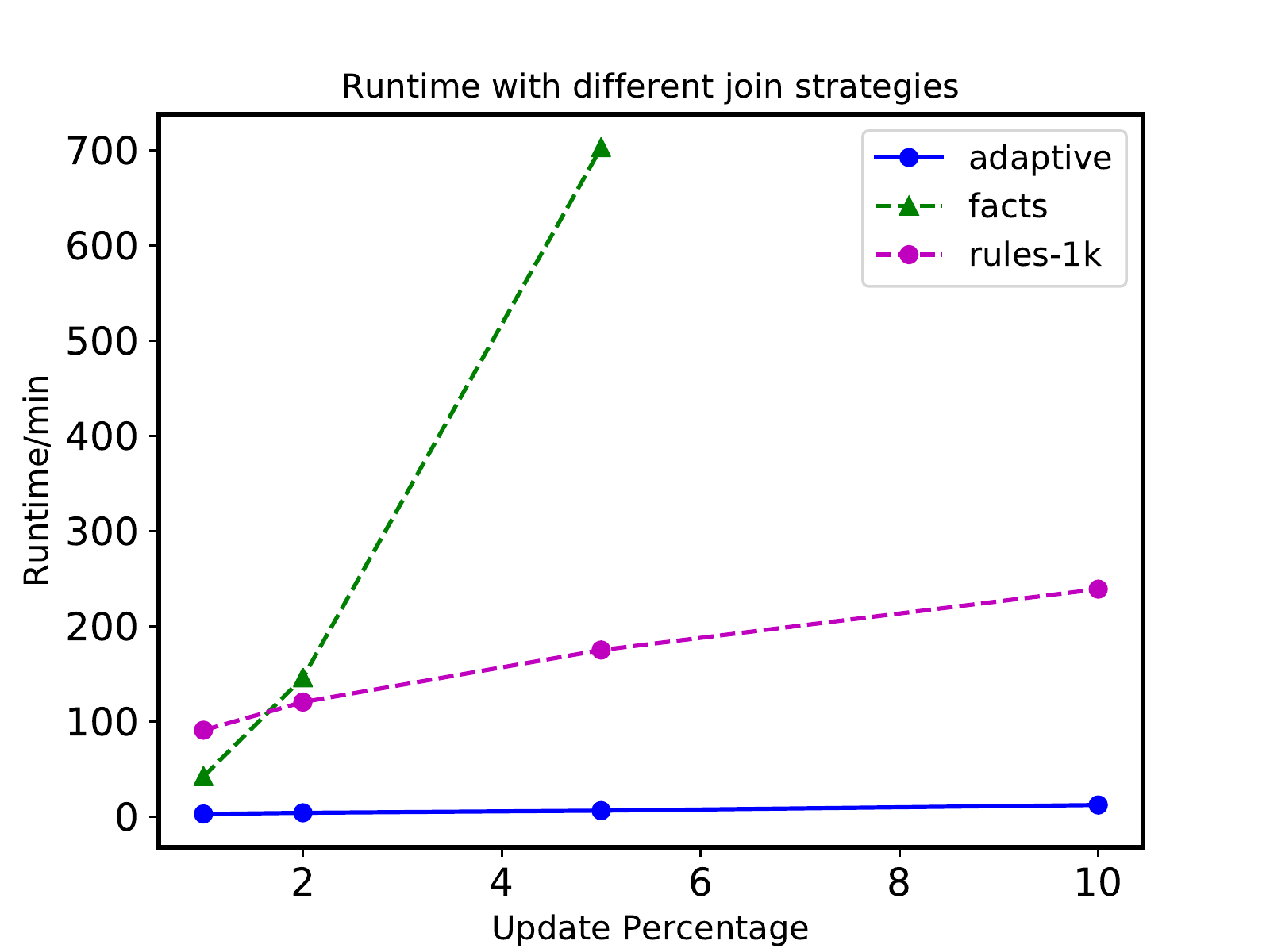}
  \caption{Different hash joins}
  \label{fig:join_diff}
\endminipage\hfill
\minipage{0.30\textwidth}
  \includegraphics[width=1.0\textwidth]{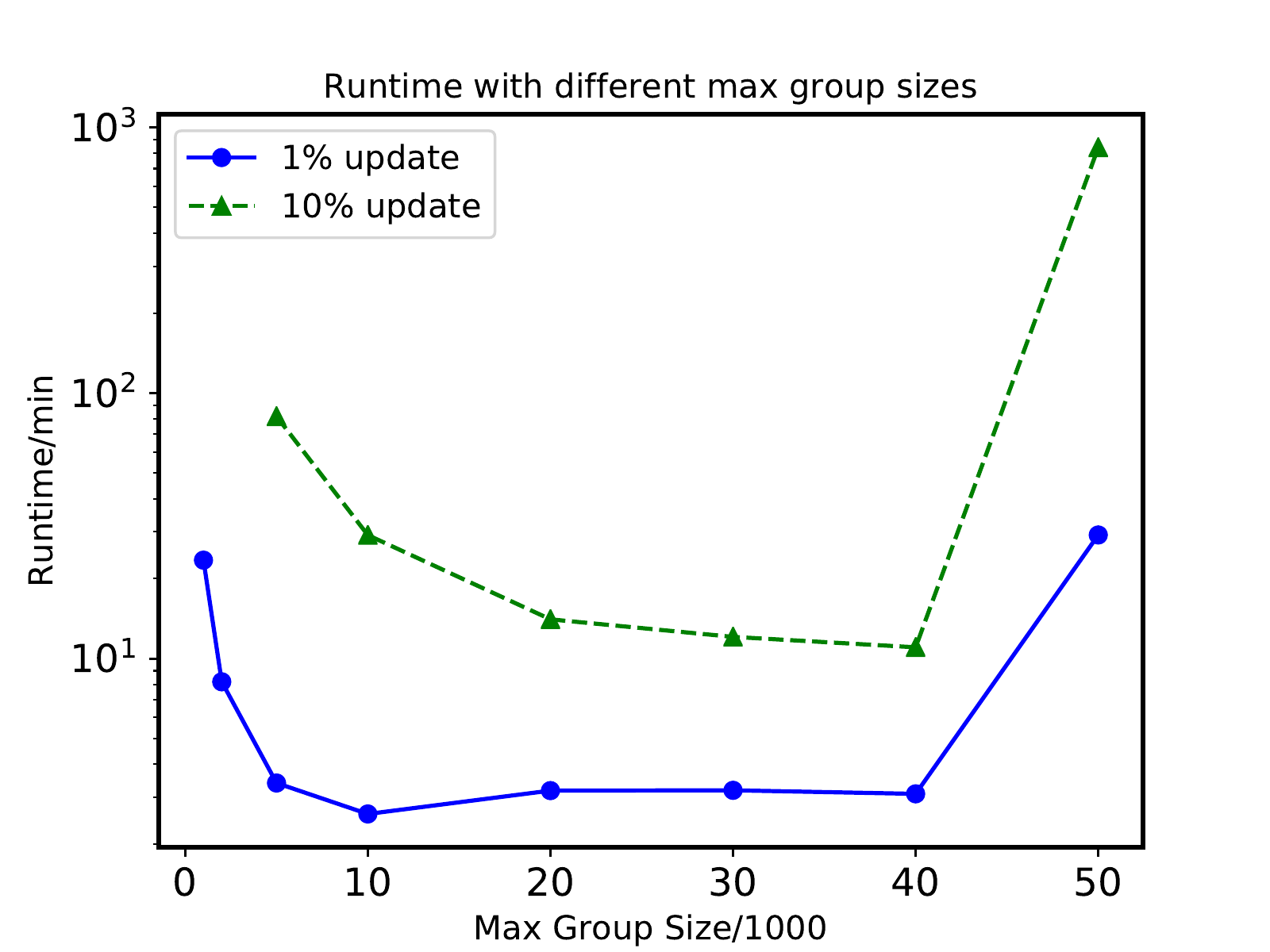}
  \caption{Different group sizes}
  \label{fig:max_group_size_diff}
\endminipage\hfill
\minipage{0.30\textwidth}%
  \includegraphics[width=1.0\textwidth]{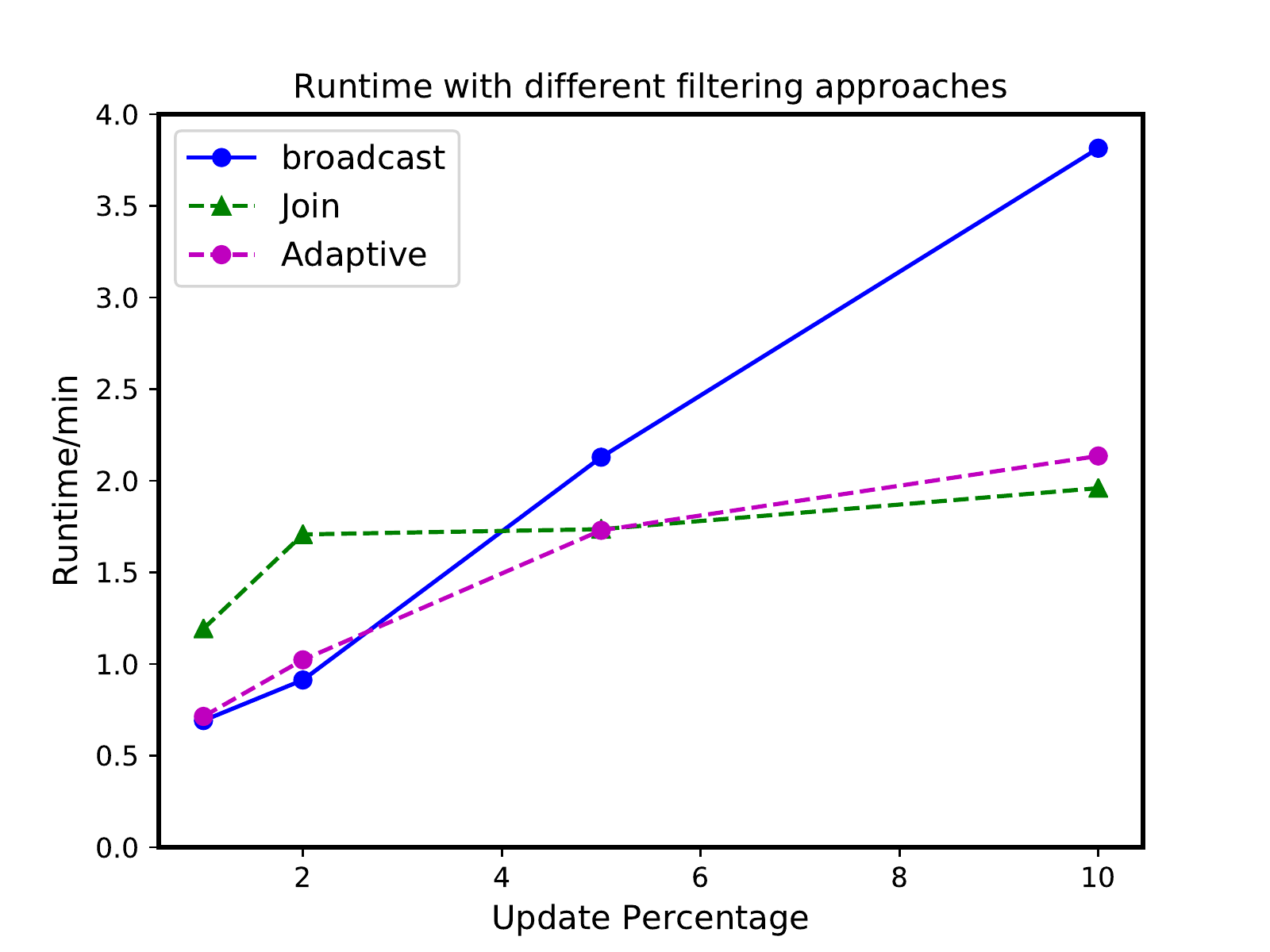}
  \caption{Different filtering}
  \label{fig:filtering_diff}
\endminipage
\end{figure*}

\textbf{Adaptive join} 
We experimentally demonstrate the effectiveness of Algorithm~\ref{alg:search}. Different searching strategies in Algorithm~\ref{alg:group-join-adaptive} lead to drastically different runtimes. Figure~\ref{fig:join_diff} shows the searching cost of 3 variants: 1) loop through the rules (same as OP) and search in facts(rules-1k); 2) loop through the paired facts and search for matching rules (facts); 3) search adaptively from the smaller size as in Algorithm~\ref{alg:group-join-adaptive} (adaptive). Variant 2 takes more than a day to finish even for a 1\% update size, thus we randomly sampled 1k rules from the 22K candidate rules, and the actual searching cost would be roughly 22 times larger. For different update sizes from 1\% to 10\%, \texttt{rules-1k} runtime reaches 700 mins at 5\% while handling only a 22th of total candidate rules; \texttt{facts} 42.2 mins to 702 mins; \texttt{adaptive} takes about 2.8 minutes to 12.3 minutes. Depending on different update sizes, \texttt{adaptive} can achieve 10x to 100x speedup. 

\para{Handle Data Skew In Groups}
Figure~\ref{fig:max_group_size_diff} shows the runtime with different group sizes to handle the data skew. For 1\% update size, the runtime drops from 23 minutes at max=1k to about 2-3 mins between max=5k to 40k and increases to 29 minutes at max=50k. Similarly, for 10\% update size, the runtime drops from 81.6 minutes at max=5k to an optimal 11 minutes at max=40k  and increases to more than 14 hours (we had to terminate instead of waiting for it to finish) at max=50. As small upper limit of the group sizes leads to more joining results, it will be a Cartesian product for all those split into small groups with the same key, while large max leads to long subtasks to finish. We choose max=30K as a default setting in our experiments.


\para{Filtering: Broadcast vs Joining}
We explore how to optimize the filtering in Algorithm~\ref{alg:search}. For a large update size $\uptau_i$, it's mandatory to use the join method because the broadcast requires $\uptau_i$ to be small enough to fit into each workers memory. Since we accommodate to various sizes of updates, we explore when to switch to the join-filtering method. Figure~\ref{fig:filtering_diff} shows the runtime of different filtering approaches on different update sizes on Freebase. \del{For broadcast filtering, the runtime increases linearly from 0.69 minutes to 3.8 minutes as update size grows from 1\% to 10\% update. For the join filtering runtime increases from 1.19 minutes to 1.96 minutes.} For smaller updates, broadcast filtering is more efficient. From the graph we set the upper size limit to use rule-based filtering to be 10M, and this achieves approximately the optimal runtime for both small(1\%) and larger(10\%) update sizes. \del{Overall, for 1\% update size, the time reduction is 0.5 mins, which reduce the runtime of Algorithm~\ref{alg:search} by about 20\% compared to join-filtering, and for 10\% update size, the reduced 1.6 mins lead to about 15\% reduction respectively.}

%% file: content/conclusion.tex
\section{Conclusion}
\label{sec:conclusion}
We propose an incremental rule mining framework that outperforms state-of-the-art batch methods on real-world datasets by more than 2 orders of magnitude. Our method leverages various optimization techniques and a new metric that significantly enhances the performance while maintaining the rule quality. We conduct experiments on two real-world knowledge bases to justify our claims. To the best of our knowledge, our system is the first that can efficiently mine rules from evolving large-scale knowledge bases with incremental updates.